\documentstyle[art10]{article}

\author{M. Temple-Raston\\
Mathematical Institute,\\
24-29 St. Giles',\\
Oxford OX1 3LB\\
United Kingdom\\
templera@maths.oxford.ac.uk}
\title{Electric monopoles in generalised $B\wedge F $ theories
}

\begin{document}

\maketitle
\begin{abstract}
A tensor product generalisation of $B\wedge F$ theories is proposed with a
Bogomol'nyi structure. Non-singular, stable, finite-energy particle-like
solutions to the Bogomol'nyi equations are studied. Unlike
Yang-Mills(-Higgs) theory, the Bogomol'nyi structure does not appear as a
perfect square in the Lagrangian. Consequently, the Bogomol'nyi energy can
be obtained in more than one way. The added flexibility permits electric
monopole solutions.

PACs: 1117,1100,0350,1210
\end{abstract}

\section{Introduction.}

It is well-known that a Bogomol'nyi structure in a Lagrangian field theory
frequently yields classical, non-singular, stable, finite-energy solutions
to the variational field equations. Moreover, these particle-like
solutions---called Bogomol'nyi solitons---appear to have far fewer quantum
corrections than might usually be expected. For example, the classical mass
spectrum for Bogomol'nyi solitons has no quantum correction; this is due to
a general relationship between supersymmetry and the Bogomol'nyi structure 
\cite{witoli}\cite{spec}. Also, for some Bogomol'nyi solitons (e.g., the BPS
magnetic monopole) the quantum corrections to the scattering differential
cross-section have been found to be remarkably and unexpectedly small \cite
{temple}.

In this letter we study a generalisation of the $B\wedge F$ topological
field theories introduced by Horowitz \cite{horowitz}. These theories are
examples of generally covariant topological gauge field theories. The
generalisation utilizes a tensor-product structure in the Lagrangian to
produce a Bogomol'nyi structure. Solitons analogous to the BPS magnetic
monopole in Yang-Mills-Higgs theory are found. Unlike Yang-Mills-Higgs
theory, however, our Lagrangian does not rely on a metric structure. The
metric structure used to define the Hodge star-operator in Yang-Mills-Higgs
theory is responsible for turning the Bogomol'nyi soliton into a magnetic
monopole. Without the metric in the tensor-product theory we are able to
construct explicit electric monopole solutions to the field equations.

\section{TFTs and Bogomol'nyi structures}

The Lagrangian field theory that forms the basis of our work is given by 
\begin{equation}
\label{TFT4} 
\begin{array}{c}
{\cal L}(A,B)=\int_{M_4}<(H^A\otimes I_E)\wedge (I_E\otimes K^B)>-\frac
12<(I_E\otimes K^B)^2> \\ +(A\leftrightarrow B,H^A\leftrightarrow K^B) 
\end{array}
\end{equation}
where $H^A$ and $K^B$ are gauge field curvatures over a four-manifold, $M_4$%
, taking values in the adjoint bundle $E$ over $M_4$. $I_E$ is the identity
transformation on the adjoint bundle, $E\rightarrow M_4$. The form of this
Lagrangian is based on the topological gauge field theories studied some
time ago by Horowitz \cite{horowitz} and the theories of Baulieu and Singer 
\cite{singer}. The four-dimensional tensor product Lagrangian gauge field
theory contains source-free electrodynamics and Yang-Mills theory \cite
{temple2}. Our interest in this letter lies in topological solitons with
rest mass. Therefore without loss of generality, we can restrict our
investigation to stationary topological solitons. This leads us to
dimensionally-reduce the four-dimensional theory using a gauge-symmetry in
time \cite{forgacs}. We denote the four-manifold quotiented by the
time-symmetry by $M_3$. Let the gauge group equal $U(n)$. $P$ is a principal 
$U(n)$-bundle over the three-manifold $M_3$. Denote the space of connections
on $P$ by ${\cal A}(P)$. We represent by $E$ the vector bundle over $M_3$
associated to $P$ by the adjoint representation. For each connection $A\in 
{\cal A}(P)$ there is an exterior covariant derivative $D^A$ acting on
sections of $E$. The covariant derivative defines a curvature $H^A$ for the
vector bundle $E$ by $D^AD^As=H^As$, where $s$ is a section of the vector
bundle $\pi :E\rightarrow M_3$. The curvature $H^A$ can be interpreted as a
2-form on $M_3$ taking values in $E$. We also introduce an equivariant Lie
algebra valued Higgs field, $\Phi _A$ , on $M_3$ paired with the connection $%
A$. In dimensional reduction the Higgs field arises as the component of the
vector potential in the direction of the gauge symmetry \cite{forgacs}. Our
starting point therefore is the energy functional dimensionally-reduced from
the Lagrangian (\ref{TFT4}). The energy functional $2\pi {\cal E}(A,B,\Phi
_A,\Phi _B)$ is given by 
\begin{equation}
\label{TFT3} 
\begin{array}{c}
\int_{M_3}<(I_E\otimes K^B)\wedge (I_E\otimes D^B\Phi _B)>\ -\
\int_{M_3}<(I_E\otimes K^B)\wedge (D^A\Phi _A\otimes I_E)> \\ 
-\int_{M_3}<(H^A\otimes I_E)\wedge (I_E\otimes D^B\Phi _B)>\ +\
(A\leftrightarrow B,H^A\leftrightarrow K^B,\Phi _A\leftrightarrow \Phi _B). 
\end{array}
\end{equation}
In the expression (\ref{TFT3}) there are two curvatures $H^A,K^B$ and two
Higgs fields $\Phi _A,\Phi _B$ corresponding to two connections $A,B\in 
{\cal A}(P)$. We assume that there is an invariant positive-definite inner
product $<\ >$ on $E_x$ which varies continuously with $x\in M_3$. The last
term in (\ref{TFT3}) symmetrises the energy functional in the dependent
fields. The energy functional is complicated, but as we shall see inherits
useful geometric structure from the four-dimensional theory presented in 
\cite{temple2}\cite{temple3}. In coordinate notation the energy functional $%
2\pi {\cal E}(A,B,\Phi _A,\Phi _B)$ can be rewritten as 
\begin{equation}
\label{TFT3'} 
\begin{array}{c}
\int_{M_3}K_{[ij}^a(D_{k]}^B\Phi _B)^b\ 
{\rm tr}(T^aT^b)\ -\ \int_{M_3}K_{[ij}^a(D_{k]}^A\Phi _A)^b\ {\rm tr}(T^aI_E)%
{\rm tr}(T^bI_E) \\ -\int_{M_3}H_{[ij}^a(D_{k]}^B\Phi _B)^b\ {\rm tr}(T^aI_E)%
{\rm tr}(T^bI_E)\ +\ (A\leftrightarrow B,\Phi _A\leftrightarrow \Phi _B). 
\end{array}
\end{equation}
Square brackets denote skew-symmetrization, and Latin subscripts run from 1
to 3. Since the gauge group is $U(n)$ we have used the Killing-Cartan form
for the bundle inner product, $<\ >$, normalised so that $<I_E{}^2>\ =\ 1$.
The variational field equations arising from (\ref{TFT3}) are 
\begin{equation}
\label{field3} 
\begin{array}{c}
D^BH^A=0,\qquad \qquad D^BD^A\Phi _A=[H^A,\Phi _B], \\ 
D^AK^B=0,\qquad \qquad D^AD^B\Phi _B=[K^B,\Phi _A]. 
\end{array}
\end{equation}

By completing the square, the energy functional (\ref{TFT3}) can be
rewritten as 
\begin{equation}
\label{top3} 
\begin{array}{c}
2\pi 
{\cal E}=\int_{M_3}<(H^A\otimes I_E-I_E\otimes K^B)\wedge (D^A\Phi _A\otimes
I_E-I_E\otimes D^B\Phi _B)> \\ -\int_{M_3}<(D^A\Phi _A\otimes I_E)\wedge
(H^A\otimes I_E)>\ +\ (A\leftrightarrow B,\Phi _A\leftrightarrow \Phi _B). 
\end{array}
\end{equation}
Let $E_A$ and $E_B$ denote the vector bundle $E$ equipped with either the
connection $A$ or $B$, respectively. We recall that the curvature of the
tensor product bundle $E_A\otimes E_B^{*}$ is given by \cite{kob}%
$$
\Omega _{E_A\otimes E_B^{*}}=H^A\otimes I_E-I_E\otimes K^B. 
$$
By defining $\Phi \equiv \Phi _A\otimes I_E-I_E\otimes \Phi _B$, then $%
D_{E_A\otimes E_B^{*}}\Phi =D^A\Phi _A\otimes I_E-I_E\otimes D^B\Phi _B$.
The first integral in (\ref{top3}) is now a topological invariant. The
Bogomol'nyi equations arising from the energy functional (\ref{top3}) are 
\begin{equation}
\label{bog} 
\begin{array}{c}
H^A\otimes I_E=I_E\otimes K^B \\ 
D^A\Phi _A\otimes I_E=I_E\otimes D^B\Phi _B 
\end{array}
\end{equation}
The first equation in (\ref{bog}) is a zero curvature condition on the
tensor product bundle $E_A\otimes E_B^{*}$. Reintroducing a coordinate
system, we can rewrite the equations in (\ref{bog}) as 
\begin{equation}
\label{bog'} 
\begin{array}{c}
H_{ij}=K_{ij}=F_{ij}(iI), \\ 
D_i^A\Phi _A=D_i^B\Phi _B=E_i(iI). 
\end{array}
\end{equation}
$F$ and $E$ are a real-valued two-form and one-form on $M_3$, respectively,
and $I$ is the identity matrix. Solutions to the Bogomol'nyi equations (\ref
{bog'}) automatically satisfy the variational field equations (\ref{field3}%
). Solutions to the first equation in (\ref{bog'}) alone are well-known to
differential geometers---they are the projectively flat connections \cite
{kob}. For line bundles projective flatness is vacuous, but for bundles of
rank greater than one projective flatness is a strong condition \cite{kob}.

Unlike the theory of BPS magnetic monopoles, the first integral in the
energy functional (\ref{top3}) vanishes with {\it either} Bogomol'nyi
equation in (\ref{bog'}) satisfied. This occurs because the first integral
in (\ref{top3}) is not in the form of a perfect square (cf.,
Yang-Mills(-Higgs) theories). With one or the other Bogomol'nyi equation
satisfied, the energy functional becomes the Bogomol'nyi energy. The
Bogomol'nyi energy is found to be 
\begin{equation}
\label{energy}2\pi {\cal E}=\int_{M_3}(D_{[k}^A\Phi _A)^aH_{ij]}^b\ {\rm tr}%
(T^aT^b)+\int_{M_3}(D_{[k}^B\Phi _B)^aK_{ij]}^b\ {\rm tr}(T^aT^b). 
\end{equation}
The extra flexibility in the topological field theory will lead us to
electric monopoles.

\section{Projectively-flat solitons}

Presumably topological monopoles, if they exist, are analogous to the BPS
magnetic monopole field configurations. Therefore we shall use the same
symmetry breaking mechanism \cite{Goddard}. The solitonic core region is
placed at the origin. Let $G$ and $G_o$ be compact and connected gauge
groups, where the group $G_o$ is assumed to be embedded in $G$. The gauge
group of the core region $G$ is spontaneously broken to $G_o$ outside of the
core region when the Higgs field is covariantly constant, $D\Phi =0$. In
regions far from the core where we assume that $D^A\Phi _A=0$, it can be
shown that 
\begin{equation}
\label{s.s.b.}H^A=\Phi _AF_A, 
\end{equation}
where $F_A\in \Lambda ^2(M_3,E_{G_o})$, a two-form on $M_3$ taking values in
the $G_o$-Lie algebra bundle, denoted by $E_{G_o}$ here \cite{Goddard}. An
equivalent expression to (\ref{s.s.b.}) can be written when $D^B\Phi _B=0$.
We shall assume that $<$$\Phi _A^2>=<\Phi _B^2>=a^2$ when $r>>1$ and where
spontaneous symmetry breaking has occurred. When $G=U(n)$ and $G_o=U(1)$, $%
F_A$ becomes a pure imaginary two-form on $M_3$. The Bogomol'nyi solitons
defined by (\ref{bog'}) have an energy (\ref{energy}) topologically fixed by 
\begin{equation}
\label{bound} 
\begin{array}{c}
2\pi 
{\cal E}=\int_{M_3}d({\rm tr}(\Phi _AH^A))+\int_{M_3}d({\rm tr}(\Phi _BK^B))
\\ =\int_{S^2}{\rm tr}(\Phi _AH^A)+\int_{S^2}{\rm tr}(\Phi _BK^B) 
\end{array}
\end{equation}
where $S^2$ is a large sphere surrounding the monopole core.

Substituting equation (\ref{s.s.b.}) into equation (\ref{bound}) and using
the asymptotic normalization condition $<$$\Phi ^2>=a^2$ for both Higgs
fields, the energy is fixed by $a^2(\int F_A+\int F_B)/2\pi $. Let us
conventionally interpret $a\int F_A/2\pi $ as the magnetic charge ($g$), and 
$a\int F_B/2\pi $ as the electric charge ($q$). We can view $F_A$ and $F_B$
as curvatures on the line bundles $L_A\rightarrow S^2$ and $L_B\rightarrow
S^2$ determined by $\Phi _A$ and $\Phi _B$, because from equation (\ref
{s.s.b.}) $F_A$ and $F_B$ are the projections of $H^A$ and $K^B$ on $L_A$
and $L_B$. The magnetic and electric charges are proportional to topological
invariants---the Chern numbers associated to complex line bundles with
curvatures $F_A$ and $F_B$, respectively. As a result both the solitonic
electric and magnetic charges are quantized at the classical level. The
Bogomol'nyi energy is given by 
\begin{equation}
\label{gq}{\cal E}=a^2(c_1(L_A)+c_1(L_B))=a(g+q). 
\end{equation}
The classical stability of the solitonic particle is also argued from the
topological invariants. Stability is assured by (\ref{gq}) if either the
electric or magnetic charge is non-vanishing.

Let us now consider non-singular, particle-like $U(n)$ solutions to {\it both%
} Bogomol'nyi equations in (\ref{bog'}) that are spontaneously broken in the
far-field. From the projective flatness of the curvatures in the Bogomol'nyi
equations (\ref{bog'}), $H^A=K^B=F(iI)$, and from equation (\ref{s.s.b.}) we
conclude that $F=\varphi _AF_A=\varphi _BF_B$, where $\Phi _A=\varphi _A(iI)$%
, $\Phi _B=\varphi _B(iI)$ and $\varphi _A,\varphi _B$ are real-valued
functions on $M_3$. The normalization of the Higgs fields implies that $%
\varphi _A^2=\varphi _B^2=a^2$. From this we find that 
\begin{equation}
\label{dyon}\int_{S^2}F_A=\pm \int_{S^2}F_B. 
\end{equation}
Therefore non-singular, stable, particle-like solutions to both Bogomol'nyi
equations (\ref{bog'}) are dyons.

To obtain electric monopoles there would appear to be two possibilities,
both resulting from a weakening of one or the other Bogomol'nyi equation (%
\ref{bog'}). We do not favour relaxing the projective flatness of the
solitons, however, because we then loose mathematical control over the nice
topological properties of the configuration space \cite{kob}. (The
configuration space is Hausdorff, presumably.) Instead, we shall maintain
projective flatness and let go of $D^A\Phi _A=D^B\Phi _B=E(iI)$, at least
asymptotically. Furthermore, since little empirical evidence exists to
suggest the independent existence of two gauge potentials, it is desirable
to restrict to a smaller set of topological solitons defined by $A=B$. We
shall call these solutions `diagonal projectively-flat solitons'. For
diagonal projectively-flat solitons the second-order variational field
equations become the Bianchi identities, and are therefore automatically
satisfied.

\section{Diagonal projectively-flat electric monopoles}

In this section we demonstrate the existence of diagonal projectively-flat $%
U(2)$ electric monopoles $(A=B,\Phi _A,\Phi _B)$ on ${\bf R}^3$ situated at
the origin. Following the example of the BPS magnetic monopole we define the
outside of the monopole to be where the gauge field is broken with a
covariantly constant Higgs field \cite{Goddard}. We assume the following
properties for the electric monopole:

\begin{enumerate}
\item  $A=B$ are projectively flat on all of ${\bf R}^3$ and take values in
the Lie algebra of $U(2)$;

\item  $A=B$ are asymptotically flat on ${\bf R}^3$;

\item  $\Phi _A,\Phi _B$ are any sufficiently differentiable functions on $%
{\bf R}^3$ taking values in the Lie algebras of $SU(n)$ and $U(n)$,
respectively;

\item  $D^B\Phi _B=0$ and $D^A\Phi _A\ne 0$, asymptotically. When $D^B\Phi
_B=0$, we assume that $\Phi _B=aI_E$ for a non-zero constant, $a$.

\item  The electric charge of the monopole is non-zero, and the magnetic
charge vanishes.
\end{enumerate}

\noindent Condition 4. is equivalent to stating that the gauge symmetry for $%
K^B$ is broken to $U(1)$ asymptotically, and that the gauge group for $H^A$
is not permitted to break far from the origin. For condition 5., assume that
a two-sphere of radius $r$, $S_r^2$, lies completely outside the monopole
that is centered at the origin. The Bogomol'nyi energy (\ref{bound}) is then
given by 
\begin{equation}
\label{case1}{\cal E}=-\frac 1{2\pi }\int_{S_r^2}F_B\ {\rm tr}(\Phi _A\Phi
_B)-\frac{a^2}{2\pi }\int_{S_r^2}F_B. 
\end{equation}
The first integral in (\ref{case1}) is the magnetic charge. The magnetic
charge vanishes since $\Phi _A$ is traceless and $\Phi _B=aI_E$, conditions
3. and 4., respectively. The second term is the topological electric charge;
the electric charge must be non-zero. We shall show that a solution
satisfying conditions 1. through 5. exists.

In general, a $U(2)$ diagonal topological soliton $(A=B,\Phi _A,\Phi _B)$
must be of the form 
\begin{equation}
\label{U2} 
\begin{array}{c}
A_j=B_j=\left( 
\begin{array}{cc}
ia_j & -c_j^{*} \\ 
c_j & ib_j 
\end{array}
\right) , \\ 
\Phi _A=\left( 
\begin{array}{cc}
i\alpha _A & -\gamma _A^{*} \\ 
\gamma _A & i\beta _A 
\end{array}
\right) ,\Phi _B=\left( 
\begin{array}{cc}
i\alpha _B & -\gamma _B^{*} \\ 
\gamma _B & i\beta _B 
\end{array}
\right) . 
\end{array}
\end{equation}
$a_j,b_j,\alpha _A,\beta _A,\alpha _B,\beta _B$ are all real-valued
functions on ${\bf R}^3$. The first Bogomol'nyi equation in (\ref{bog'})
states that the vector potential is projectively flat, $H^A=K^B=F(iI)$. A
straightforward calculation of $H^A$ informs us that 
\begin{equation}
\label{Fij} 
\begin{array}{ccc}
F_{ij} & = & \partial _ia_j-\partial _ja_i+i(c_j^{*}c_i-c_i^{*}c_j), \\  
& = & \partial _ib_j-\partial _jb_i-i(c_j^{*}c_i-c_i^{*}c_j). 
\end{array}
\end{equation}
and that 
\begin{equation}
\label{comp}\partial _ic_j-\partial _jc_i-i\left[
(c_ia_j-c_ja_i)-(c_ib_j-c_jb_i)\right] =0. 
\end{equation}
Equation (\ref{Fij}) in coordinate free notation becomes 
\begin{equation}
\label{Fij'}F=d{\bf a}+i{\bf c}^{*}\wedge {\bf c}=d{\bf b}-i{\bf c}%
^{*}\wedge {\bf c}, 
\end{equation}
and implies that ${\bf c}^{*}\wedge {\bf c}=2id({\bf b-a})$, that is, ${\bf c%
}^{*}\wedge {\bf c}$ is exact. Similarly, equation (\ref{comp}) can be
rewritten as 
\begin{equation}
\label{comp'}d{\bf c}+i({\bf a}-{\bf b})\wedge {\bf c}={\bf 0}, 
\end{equation}
where ${\bf a}=a_i\ dx^i$, ${\bf b}=b_i\ dx^i$, and ${\bf c}=c_i\ dx^i$. We
must now write down a solution to equations (\ref{Fij'}) and (\ref{comp'})
that can be shown latter to have non-vanishing electric charge.

We introduce the complex coordinate $\zeta ={\bf P}_r(r,\theta ,\varphi )$
that comes from the stereographic projection, ${\bf P}_r$, of the spherical
polar coordinate $(r,\theta ,\varphi )$ on the two-sphere minus the north
pole, $S_r^2-\left\{ N\right\} $, to the complex plane minus inifinity, $%
{\bf CP}^1\backslash \left\{ \infty \right\} $. The projected 1-forms ${\bf a%
}$ and ${\bf b}$ on ${\bf CP}^1\backslash \left\{ \infty \right\} $ will
also be denoted by ${\bf a}$ and ${\bf b}$. Assume that ${\bf a}$ and ${\bf b%
}$ on ${\bf CP}^1\backslash \left\{ \infty \right\} $ differ by the
non-exact real-valued 1-form, $\chi $; that is, ${\bf b}={\bf a}+\chi $. Now
define 
\begin{equation}
\label{csolu}{\bf c}(\zeta )=i\sqrt{2}\left[ \exp i\int_{{\bf P}\gamma }(%
{\bf b}-{\bf a})\right] \ d\zeta ,
\end{equation}
where the contour integration is along the stereographic projection of the
great circle, $\gamma $, from the south pole of $S_r^2$ to the spherical
polar coordinate given by $(r,\theta ,\varphi )={\bf P}_r^{-1}(\zeta )$.
Assume that both the 1-forms ${\bf a}$ and ${\bf b}$ vanish at the south
pole for all values of $r$. It is easy to verify that (\ref{csolu})
satisfies equation (\ref{comp'}), and that ${\bf c}^{*}\wedge {\bf c}=2i\
d\chi \equiv -2d\bar \zeta \wedge d\zeta $.

Turning to the Higgs fields, recall that we require that $A$ be $su(2)$%
-valued everywhere, and that the gauge group for $B$ is broken to $U(1)$
asymptotically ($D^B\Phi _B\rightarrow 0$ as $r\rightarrow \infty $). Since
the Higgs field $\Phi _A$ takes values in the Lie algebra of $SU(2)$, then $%
\parallel \gamma _A\parallel ^2=1+\alpha _A\beta _A$. To demonstrate the
existence of an electric monopole, take $\alpha _A=\beta _A=0$ so that $%
\parallel \gamma _A\parallel ^2=1$. For $\Phi _B$, the only condition
imposed on the Higgs fields is that $\Phi _B=aI_E$, asymptotically.
Substitute (\ref{U2}) into the asymptotic equation $D^B\Phi _B=0$ to find
that on the diagonal 
\begin{equation}
\label{diag}d(\alpha _B+\beta _B)=0,\qquad d\alpha _B=i\left( {\bf c}\gamma
_B^{*}-{\bf c}^{*}\gamma _B\right) , 
\end{equation}
and off the diagonal, 
\begin{equation}
\label{offdiag}d\gamma _B=i\left( \left( \beta _B-\alpha _B\right) {\bf c}%
+\left( {\bf a}-{\bf b}\right) \gamma _B\right) ,\qquad {\rm complex\ conj.} 
\end{equation}
For $\left| \zeta \right| \gg 1$, let%
$$
\alpha _B(\zeta ,\bar \zeta )=a+\frac 1{\left( 1+\left| \zeta \right|
^2\right) ^2}+O((\bar \zeta \zeta )^{-3}),\qquad 
$$
$$
\beta _B(\zeta ,\bar \zeta )=a+\frac 1{\left( 1+\left| \zeta \right|
^2\right) ^2}+O((\bar \zeta \zeta )^{-3}). 
$$
$\alpha _B=\beta _B=a$ asymptotically is consistent with the requirement
that $\Phi _B=aI_E$, and satisfies the asymptotic equations (\ref{diag}) and
(\ref{offdiag}). The remaining asymptotic equations for $\gamma _B$ become%
$$
\begin{array}{ccc}
c_i\gamma _B^{*}-c_i^{*}\gamma _B & = & 0, \\ 
\partial _i\gamma _B+i\gamma _B(b_i-a_i) & = & 0. 
\end{array}
$$
$\gamma _B=0$ is a solution to these equations and is also consistent with
the requirement that $\Phi _B=aI_E$.

Now we compute the electric charge. The broken gauge far-fields are given by 
\begin{equation}
\label{Fbroken}
\begin{array}{ccccccc}
F_A & \equiv  & <H^A\Phi _A> & = & -iF{\rm tr}(\Phi _A) & = & 0, \\ 
F_B & \equiv  & <K^B\Phi _B> & = & -iF{\rm tr}(\Phi _B) & = & F(\alpha
_B+\beta _B)/2,
\end{array}
\end{equation}
Recall that the magnetic charge of the solution, given by $\lim
{}_{r\rightarrow \infty }a\int F_A/2\pi $, vanishes because $\Phi _A$ is
traceless. The solitonic electric charge is given by%
$$
2\pi q=a\lim {}_{r\rightarrow \infty }\int_{S_r^2}F_B=-2ia\int_{{\bf CP}^1}
\frac{d\bar \zeta \wedge d\zeta }{\left( 1+\left| \zeta \right| ^2\right) ^2}%
=-4\pi a,\  
$$
where ${\bf a}$ is closed, but ${\bf b}$ is not closed. The sphere $S_r^2$
is centered round the monopole at the origin and is assumed to lie
completely in regions where the gauge field has been broken. Notice that the
electric charge within $S_r^2$ depends only on the gauge potential, and is
time-independent, gauge-invariant, and unchanged under any continuous
deformation of the enclosing surface. This proves the existence of an
electric monopole within this theory.

To interpret these solitons, we note that $U(2)$ is double covered by $%
SU(2)\times U(1)$. Also, the classical mass and particle spectrum of the
projectively-flat electric monopole compare favourably with that of the
intermediate vector bosons. The mass of the electric monopole is $M=aq$, the
same as the mass of the $W^{\pm }$. Moreover, there are no quantum
corrections to the classical mass, because of the general relationship
between supersymmetry and the Bogomol'nyi structure \cite{witoli}\cite{spec}%
. The $Z_0$, presumably, corresponds to the case where $H^A=K^B=FI_E$, but $%
D^A\Phi _A\neq 0$ and $D^B\Phi _B\neq 0$, that is, there is no symmetry
breaking. The gauge far-fields in that case are non-abelian and pass
unnoticed through the detectors. Although uncharged the soliton's energy is
topologically fixed by%
$$
\begin{array}{ccc}
2\pi {\cal E} & = & \int_{S^2} 
{\rm tr}(\Phi _AH^A)+\int_{S^2}{\rm tr}(\Phi _BK^B) \\  & = & 
-\int_{S^2}F<\Phi _A>-\int_{S^2}F<\Phi _B> 
\end{array}
$$
So it, too, is stable under perturbations. We propose therefore that the
solitons in the tensor product topological field theory defined in section
two form a provisional model for the $W^{\pm }$ and $Z_0$ intermediate
vector bosons. Many years ago it was promoted that intermediate vector
bosons should appear as Bogomol'nyi solitons dual in some sense to the BPS
magnetic monopole \cite{olive}.

\section{Conclusion.}

The tensor product energy functional (\ref{TFT3}) has a Bogomol'nyi
structure and solitonic particle solutions. The projective-flatness observed
in the first Bogomol'nyi equation (\ref{bog'}) is well-known to be related
to algebraic stability in complex vector bundles, and used in the
construction of well-behaved moduli spaces \cite{kob}. Moduli spaces in this
context are often seen to be the covariant phase spaces and configuration
spaces of the physical theory, so it is not surprising to uncover a
requirement for projective-flatness, although such a requirement is not
necessary. As the Bogomol'nyi equations do not arise from a perfect square,
there is more flexibility in achieving the Bogomol'nyi energy. An
interesting class of soliton has been studied herewithin by restricting to
those solutions of the Bogomol'nyi equations where the gauge potentials are
equal, $A=B$. When both Bogomol'nyi equations are satisfied stable,
particle-like solitons are dyonic. By relaxing the Bogomol'nyi structure
stable, particle-like solutions can be found that carry only an electric
charge. Similarities exist between the heavy intermediate vector bosons in
the standard model and the solitons in this theory.

\end{document}